\documentclass[apjl]{emulateapj}
\usepackage{natbib}
\usepackage{enumerate}
\usepackage{color}
\usepackage{xfrac}
\usepackage[bottom]{footmisc}
\usepackage{subfigure}
\usepackage{amsmath}
\usepackage{graphicx}

\newcommand{\Al}{$\,^{26}$Al}
\newcommand{\be}{\begin{equation}}
\newcommand{\ee}{\end{equation}}
\newcommand{\ba}{\begin{align}}
\newcommand{\ea}{\end{align}}
\newcommand{\shat}{ {\hat s} } 
 
\newcommand{\mbar}{ \langle m \rangle} 
\newcommand{\erf}{ {\rm erf} } 
 
\newcommand{\edense}{ {\cal E} }

\begin{document}

\title{Radionuclide Ionization in Protoplanetary Disks: Calculations of Decay Product Radiative Transfer} 

\author{L. Ilsedore Cleeves$^1$, Fred C. Adams$^{1,2}$, 
Edwin A. Bergin$^1$, Ruud Visser$^1$}

\affil{$^1$Department of Astronomy, University of Michigan, 
500 Church Street, Ann Arbor, MI 48109}

\affil{$^2$Department of Physics, University of Michigan, 
450 Church Street, Ann Arbor, MI 48109}

\begin{abstract}
We present simple analytic solutions for
the ionization rate $\zeta_{\rm{SLR}}$ arising from the decay of
short-lived radionuclides (SLRs) within protoplanetary disks. We solve
the radiative transfer problem for the decay products within the disk,
and thereby allow for the loss of radiation at low disk surface
densities; energy loss becomes important outside $R\gtrsim30$~AU for
typical disk masses $M_g=0.04$~M$_\odot$.  Previous studies of
chemistry/physics in these disks have neglected the impact of
ionization by SLRs, and often consider only cosmic rays (CRs), because
of the high CR-rate present in the ISM.  However, recent work suggests
that the flux of CRs present in the circumstellar environment could be
substantially reduced by relatively modest stellar winds, resulting in
severely modulated CR ionization rates, $\zeta_{\rm{CR}}$, equal to or
substantially below that of SLRs ($\zeta_{\rm{SLR}}\lesssim10^{-18}$~s$^{-1}$). 
We compute the net ionizing particle fluxes and
corresponding ionization rates as a function of position within the
disk for a variety of disk models. The resulting expressions are
especially simple for the case of vertically gaussian disks
(frequently assumed in the literature).  Finally, we provide a
power-law fit to the ionization rate in the midplane as a function of
gas disk surface density and time. Depending on location in the disk,
the ionization rates by SLRs are typically in the range
$\zeta_{\rm{SLR}}\sim(1-10)\times10^{-19}$~s$^{-1}$.
\end{abstract}

\keywords{accretion, accretion disks --- circumstellar matter --- radiative transfer --- stars: pre-main sequence} 

\section{Introduction} 
Ionization plays an important role in setting thermal, dynamical, and
chemical properties of protoplanetary disks.  The dominant ionization
processes thought to be active in such disks include photoionization from stellar
and interstellar UV and X-ray radiation, thermal ionization,
ionization by the decay products of short-lived radionuclides (SLRs),
and cosmic ray (CR) ionization
\citep[e.g.,][]{glassgold1997,finocchi1997,glassgold2001,walsh2012}.
Of these sources, only CR and SLR-decay are able to provide ionization
in the densest and coldest layers of the disk where UV and X-ray photons are
highly attenuated. However, the importance of CR ionization is highly
uncertain, and position dependent, due to stellar-wind modulation from
the central star \citep[][hereafter Paper I]{cleeves2013}. With the
substantially reduced CR rates expected for these disk systems, SLR-decay is left as the dominant midplane ionization contributor at
distances beyond the hard X-ray dominated region, $R\gtrsim10$~AU from the central star.

Indeed, the Solar System's meteoritic record points to an early
enhancement \citep[$\sim10$ times the mean ISM abundance;][]{umebayashi2009} of \Al, the most chemically
significant of the SLRs, indicating an enrichment of massive star
byproducts in the Solar birth cluster \citep{adams2010}. Furthermore,
maps of 1.808 MeV $\gamma$-rays resulting from \Al\ decay confirm an
enhancement of SLRs near major star-forming regions
\citep{diehl2006}.  While the ubiquitous presence of SLRs during
star-formation and subsequent disk-formation is expected, the degree
to which their energetic decay products contribute to the ionization
rate remains uncertain.  For example, there is inherent time evolution
in both the total mass of SLRs \citep[set by their respective half-lives where $t=0$ corresponds to the time of formation of CAIs;][]{macpherson1995}
and the spatial distribution of dust particles that carry the SLRs, which tend
to settle towards the midplane with time
\citep{umebayashi2013}. Furthermore, the diversity of original sources
of radioactive particles, including supernovae \citep{cam77},
Wolf-Rayet winds \citep{arnould1997,gaidos2009}, and stellar
spallation \citep{lee1998,shang2000}, adds further complexity to
characterizing the initial abundances of SLRs present at the time of
disk-formation \citep[see also][]{adams2010}. Nevertheless, the
meteoritic record provides clues regarding which species were once
present in one particular protoplanetary disk, our Solar Nebula, and
the abundances therein \citep[e.g.,][]{gounelle2012}.

A number of studies have quantified the ionization of molecular gas by
energetic particles resulting from SLR decay
\citep[e.g.,][]{umebayashi1981,finocchi1997,umebayashi2009}.  For a
disk with an incident CR flux at ISM levels, previous work predicts
that ionization by CRs should exceed that of SLRs in regions where the
gas surface density $\Sigma_g<1000$~g~cm$^{-2}$
\citep{umebayashi1981,umebayashi2009}. However, under the influence of
a wind-reduced CR flux (Paper~I), even a modest ``present-day'' solar
wind reduces the CR ionization rate to values rivaling or
substantially below that of SLRs ($\zeta_{\rm{CR}}\lesssim10^{-18}$
s$^{-1}$) {\em{throughout the disk}}. Careful treatment of ionization
by SLRs is thus necessary for models of disk chemistry and physics.
Furthermore, previous studies of disk ionization by SLRs did not take
into account the escape of the decay products, which becomes important
when the surface density drops below
$\Sigma_g\lesssim10~{\rm{g~cm^{-2}}}$.

In the present paper we develop easy-to-use approximations to implement
position-dependent SLR ionization rates in protoplanetary disks,
$\zeta_{\rm{SLR}}(r,z)$, for use in chemical models and/or studies of
the magnetorotational instability \citep{balbus1991}.
Section~\ref{sec:slab} calculates the ionization rate using a
plane-parallel approximation for the radiative transfer, which
explicitly includes the escape of decay products from the disk.
Although this paper focuses on the dominant ionizing source, \Al, we
examine the effects of including other SLR species, $^{60}$Fe and
$^{36}$Cl, on the estimated ionization rates. Section~\ref{sec:settle}
generalizes the calculation to include the effects of dust settling on
the ionization rates.  Finally, Section~\ref{sec:time} considers the
time evolution of the ionization rates.

\section{Transfer of Short-Lived Radionuclide Decay Products}\label{sec:radtran}

For a given parent SLR, the decay process can result in the emission
of $E\sim1$~MeV photons, positrons, electrons and $\alpha$-particles,
whose energy goes into ionization, excitation, and heating of the
surrounding gas via secondary electrons 
\citep[see][]{umebayashi1981,dalgarno1999,glassgold2012}. Because the
SLR mass reservoir in the disk is finite, the half-life
$t_{\rm{half}}$, in addition to the mass/abundance of parent SLRs, is
vital, as it sets both the total duration and occurrence rate of the
decays.  A short half-life results in frequent decays (high ionization
rates), but lasts for a potentially negligible fraction of the disk
lifetime.  From the ionizing secondary electrons, the energy required to
create a single ion-pair from H$_2$ gas is $W_{\rm{H_2}}=36$~eV, where
only $\sim$47\% of the energy goes into ionization
\citep{dalgarno1999,glassgold2012}. As the decay products propagate
through the gas disk, the main source of opacity for $E\sim1$~MeV
photons is Compton scattering and for positrons/electrons is
collisional ionization. As a result, the decay products have finite,
energy-dependent ranges. The branching ratios, ranges and decay
sequences of the SLR parents considered in this work are shown in
Table~\ref{tab:nucdat}. In the following section we compute H$_2$
ionization rates; however, these results can be extended to include
helium ionization, where $\zeta_{\rm{He}}\approx0.84\zeta_{\rm{H_2}}$
\citep{umebayashi2009}.
\begin{deluxetable}{ccccc}
\begin{centering}
\tablecolumns{4} 
\tablecaption{Selected SLR Data. \label{tab:nucdat}}
\end{centering}
%\tablewidth{3pt}
\tabletypesize{\footnotesize}
\tablehead{Parent SLR &$t_{\rm{half}}$&Decay& Product &$\kappa$ \\ 
 log$_{10}$($n_\chi/n_{\rm{H_2}}$)&(Myr)&Mode&(MeV)&(cm$^{2}$/g) }
\startdata
\Al\ (-9.378)&0.74& $\beta^+$ 82\%  & e$^+$ (0.473) &11.76\\
&&& $\gamma$ ($2\times0.511$) &0.148\\
 &&& $\gamma$ (1.808) &0.080\\
 && E.C.\footnote{Electron capture.} 18\% & $\gamma$ (1.808)&0.080\\
$^{60}$Fe (-10.270)&1.5&$\beta^-\left[{\rm{^{60}Co}}\right]$&e$^-$(0.184) & 52.63 \\
&&& $\gamma$ (0.0586) &0.282\\
 &&$\beta^-\left[{\rm{^{60}Ni}}\right]$\footnote{The ${\rm{^{60}Co}}$ half-life is 5.3yr, and therefore we assume it happens instantaneously after the $^{60}$Fe decay event.}&e$^-$(0.315) &21.74\\
 && & $\gamma$ (1.173)&0.101\\
 && & $\gamma$ (1.332)&0.094\\
$^{36}$Cl (-10.367)&0.30&$\beta^-$\footnote{There are formally two decay modes: $\beta^{-}$ (98\%) and $\beta^{+}$ (2\%). We include only the former in our calculations.}&e$^-$(0.7093)& 7.541
\enddata
\end{deluxetable}

The problem of interest is essentially a classical radiative transfer
problem where the distribution of emitters follows the refractory
material (assuming that most of the radioactive metals are carried by
dust grains). Furthermore, we assume that the decay products escape
the dust grains from which they are emitted, which is appropriate for
dust grains sizes $a\le10$~cm for $E\sim1$ MeV photons and
$a\le0.1$~cm for positrons \citep{umebayashi2013}.  The absorption is
due to energy losses in the gas, and the resulting equation for the
frequency-averaged particle/photon intensity, $I$, has the general form  
\be\label{eq:basic}
\shat\cdot\nabla{I}=\rho_g\left({100\over{f_g}}\right){J\over4\pi}-\rho_g\kappa{I},\ee 
where $J$ is the emissivity from the production of
photons/positrons/electrons due to radioactive decay, $\kappa$ is the
mass absorption coefficient, $\rho_g$ is the gas density, and $f_g$ is
the gas-to-dust mass fraction ($f_g=100$ for uniformly distributed 
gas and dust).

To start, we neglect energy/frequency evolution of the decay product
cross sections (see Section~\ref{sec:discussion}). To specify $J$, we
consider decay product ($k$) from a single radioactive species
($p$).  If $E_k$ is the energy of the decay product and
$\omega_p=\log{2}/t_{\rm{half}}$ is the decay rate, then $E_k\omega_p$
is the energy generated by $k$ per second per parent SLR $p$. Let
$\chi_p$ be the abundance of the parent species relative to H$_2$
and let $\mbar=\mu{m_{\rm{H}}}$ denote the mean molecular
weight of gas (where $\mu\approx2.36$). We thus expect the emissivity
$J_k$ to have the form
\be\label{eq:emissivity}J_k={E_k\omega_p\chi_p\over\mbar}\,\,\left[{\rm{erg}}\,{\rm{s}}^{-1}{\rm{g}}^{-1}\right].\ee
In the regime where all decay products are trapped, the ionization
rate per H$_2$ is given by the usual expression
$\zeta_{\rm{H_2}}^k=E_k\omega_p\chi_p/W_{\rm{H_2}}~{\rm s^{-1}}$
\citep[e.g.,][]{umebayashi1981}. This exercise can be extended over
all parent SLRs $p$, and decay products $k$, within a decay series,
thus obtaining:
\be\label{eq:sum}\zeta_{\rm{H_2}}^{\rm{tot}}(r,z)=\sum_p\sum_k\zeta_{\rm{H_2}}^{p,k}(r,z)\,.\ee

While the following section provides ionization prescriptions that apply to
general disk models, the plots are calculated using our model
from Paper I.  The gas and dust densities are described by a power-law
with an exponential taper at the outer-edge \citep{andrews2011} with a total disk gas mass of $M=0.04~{\rm{M_\odot}}$.
Formally, the vertical dust profile is described by two gaussians for
millimeter- (midplane) and micron-sized (atmospheric) grains.  For our
well-mixed calculation in Section~\ref{sec:slab}, we assume instead
that the dust follows the gas with a uniform gas-to-dust mass ratio
$f_g=100$. For the settled disk discussed in Section~\ref{sec:settle},
we compare different methods by which to approximate
$\zeta_{\rm{SLR}}$ with more sophisticated dust profiles. 

\subsection{Plane-Parallel Approximation}\label{sec:slab}

\begin{figure}\begin{centering}
\includegraphics[width=3.3in]{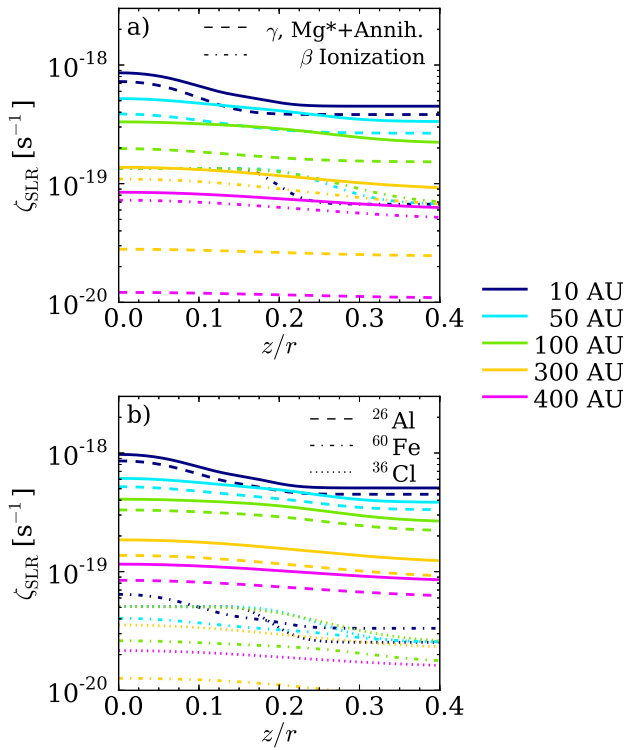}
\caption{H$_2$ ionization rate as a function of normalized height
$(z/r)$ and at specified disk radii
(indicated by line color).  {\em{Top panel:}} \Al\ ionization broken   
into photon (dashed) and particle processes (dot-dash); solid line
represents the total ionization rate. {\em{Bottom panel:}} Total 
ionization from \Al,
$^{60}$Fe, and $^{36}$Cl; individual contributions indicated by dashed, dot-dash, and dotted
lines, respectively. \label{fig:ZetaAllMix}}
\end{centering}\end{figure}

For disks where the radial variation in density is much slower than
variations with height $z$, we can treat the disk as essentially an
``infinite slab.''  We thus carry out a plane-parallel radiative
transfer calculation where intensity $I$ is a function of height $z$
above the midplane, and one direction variable. The ray direction is
determined by the angle $\theta$ with respect to the $z$-direction,
or, equivalently, $\mu=\cos\theta$. The validity of this approximation 
requires that the disk scale height ${H}\ll{r}$. 

In the well-mixed case, the radiative transfer equation (\ref{eq:basic}) 
has the formal solution:
\begin{align}\label{formsolution}I(z,\mu)&={1\over4\pi}\int_0^\infty\rho{J}\exp[-\tau(z,\mu;s)]ds
\nonumber\\&={1\over4\pi}\int_0^{\tau_\mu(z)}{J\over\kappa}\exp[-\tau(z,\mu;s)]d\tau,\end{align}
where the optical depth along a ray in the $\mu$-direction is 
\be{\tau(z,\mu;s)=\int_{0}^{s}\kappa\rho(s'){ds'}.}\label{opdep}\ee 
Equation~(\ref{formsolution}) allows both the emissivity $J$ and the
opacity $\kappa$ to vary with position. In many applications, however,
these quantities are constant.  In general, we want to evaluate the
intensity $I$ at a given height $z$ and for a given ray direction
specified by $\mu$. Along the ray, the cylindrical height $z'$ is
given by \be{z'=z-\mu{s}\,.}\ee Given the optical depth expression
(\ref{opdep}), and the substitution $dz'=-\mu{ds}$, we can write the
optical depth $\tau_\mu(z)$ in the $\mu$-direction in terms of the
vertical optical depth $\tau_{\pm}$,
\be\tau_\mu(z)=-{1\over\mu}\int_{z}^{\pm\infty}\kappa\rho(z'){dz'}\equiv{1\over|\mu|}\tau_{\pm}(z)\,.\ee
By separating out the angular dependence from the optical depth, we
only need to calculate the $\tau_{\pm}$ integrals once for a given
location.  Accordingly, the specific intensity is given by 
\begin{align}\label{eq:mixinten}I(z,\mu)=&{J\over{4\pi\kappa}}\left(1-\exp{\bigl[-\tau_\mu(z)\bigr]}\right)\nonumber\\=&{J\over{4\pi\kappa}}\left(1-\exp{\left[-{1\over|\mu|}\tau_\pm(z)\right]}\right)\,.\end{align}
We integrate (\ref{eq:mixinten}) over solid angle to determine the total energy
density in ionizing decay photons/particles $\edense(z)$:
\begin{align}c\edense(z)&=\int{I}(z,\mu){d}\Omega=2\pi\int_{-1}^{1}{I}(z,\mu){d}\mu\,\nonumber\\&=2\pi\int_{-1}^{1}{J\over{4\pi\kappa}}\left(1-\exp{\left[-{1\over\mu}\tau_\pm(z)\right]}\right){d}\mu.\end{align}
Substituting $t=1/|\mu|$, we can evaluate this expression
in terms of exponential integrals of order two \citep{abramowitz1972},
\begin{align}\label{eq:formEnD}c\edense(z)=&{J\over{2\kappa}}\left\{2-\int_{1}^{\infty}{1\over{t^2}}\exp{\left[-t\tau_+(z)\right]}{d}t\right.\nonumber\\&-\left.\int_{1}^{\infty}{1\over{t^2}}\exp{\left[-t\tau_-(z)\right]}{d}t\right\}\nonumber\\=&{J\over{2\kappa}}\left\{2-E_2\left[\tau_+(z)\right]-E_2\left[\tau_-(z)\right]\right\}.\end{align}
The ionization rate is then given by 
\begin{align}\zeta_{\rm{H_2}}^k(z)&={c\edense(z)\mbar\kappa\over{W_{\rm{H_2}}}}\label{eq:zetagen}\\&={1\over2}{{E_k\omega_p\chi_p}\over{W_{\rm{H_2}}}}\left\{2-E_2\left[\tau_+(z)\right]-E_2\left[\tau_-(z)\right]\right\}\label{eq:zeta_mix}\end{align}
Equation~(\ref{eq:zeta_mix}) provides the general solution for the
ionization rate due to decay product $k$ for a well-mixed disk. This
expression can be readily evaluated with knowledge of the disk's
opacity in the $z$-direction. Note that the functions $E_2(z)$
are standard (e.g., \citealt{abramowitz1972}, or the Python
library {\em{SciPy}}).

\begin{figure*}
\begin{centering}
\includegraphics[width=6.8in]{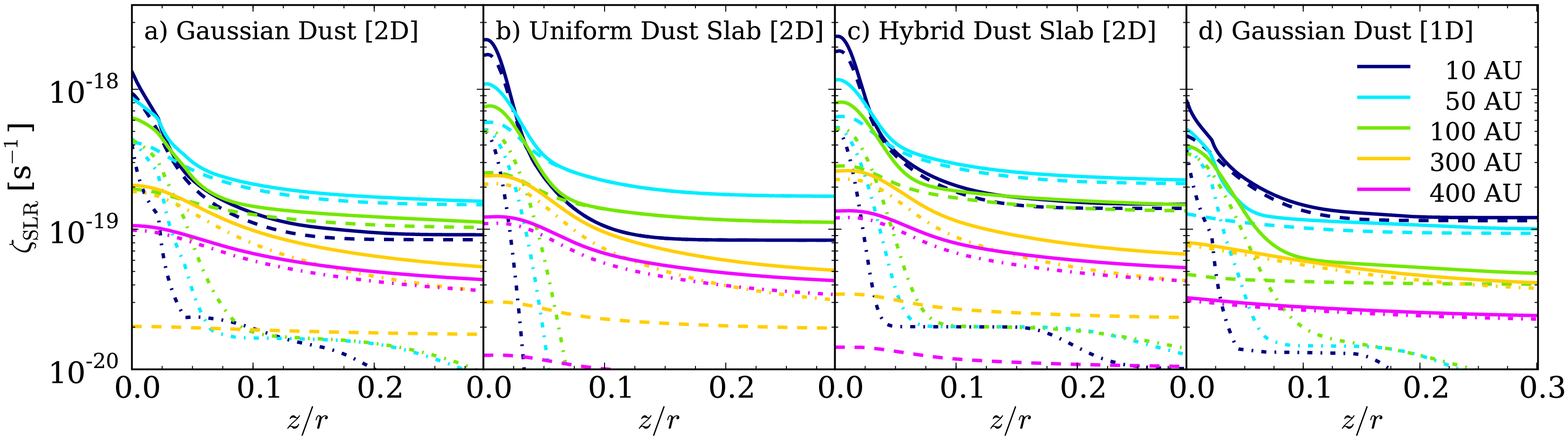}
\caption{Line-styles as indicated in Figure~\ref{fig:ZetaAllMix} for \Al\ ionization only:  
a) infinite slab calculation for the original model dust profile; b)
infinite slab calculation for a uniform density dust layer in the
midplane; c) hybrid model of uniform dust layer combined with a
small-grain gaussian distribution; and d) original density profile
with decay product escape restricted to the $z$-direction. \label{fig:ZetaSettle}}
\end{centering}
\end{figure*}

For a vertically gaussian disk of the form
$\rho_g(z)=\rho_0\exp{[-1/2~(z/H)^2]}$, we can solve exactly for $\tau_\pm$,
\be\label{eq:tauformal}\tau_\pm(z)=\tau_0\left\{1\mp{\rm{erf}}\left[{z\over\sqrt{2}H}\right]\right\},\ee
where we have defined
\be\label{eq:tauzero}\tau_0=\kappa\rho_0H\sqrt{\frac{\pi}{2}}={\kappa\Sigma_g\over2}.\ee
Note that $\tau_\pm$ must be computed for each type of decay product
$k$ where the decay ranges are provided in Table~\ref{tab:nucdat}.
Nonetheless, Equation~(\ref{eq:tauformal}) allows the ionization rate
to be calculated -- easily and exactly -- from
Equation~(\ref{eq:zeta_mix}) at any point $(r,z)$ in the disk.

Figure~\ref{fig:ZetaAllMix} presents the results from
Equations~(\ref{eq:zeta_mix},\ref{eq:tauformal}) for our
standard disk model. Of the various SLR parent bodies, \Al\ dominates
the ionization rate $\zeta_{\rm{SLR}}$ at early times, as demonstrated
in the bottom panel where we include ionization contributions from
\Al, $^{60}$Fe, and $^{36}$Cl decay, which together increase
$\zeta_{\rm{SLR}}$ by $\sim13$\%.  Among the \Al\ decay products,
within $R<100$~AU the more energetic $\gamma$-rays play the largest
role, while outside of this region the more readily stopped positrons
carry the ionization (see Figure~\ref{fig:ZetaAllMix}, dashed and
dot-dash lines, respectively).

\subsection{Dust Settling}\label{sec:settle}

As dust settles towards the midplane, the sources of emissivity (SLRs
in the dust) and the absorbers (gas) are no longer well-mixed. In our
settled disk model (Paper I), the large grains have a smaller
scale-height $h$ than the gas and small grains, $H$, where both dust
reservoirs are described by a gaussian profile and where $h/H=0.2$.
This section provides two methods to approximate this more complicated
radiative transfer problem.

The original integral for the energy density can be written in the form
\begin{align}\label{eq:xisettle}c\edense(z)={J\over2}\int_{-1}^{1}d\mu\int_{0}^{\infty}\rho_g(s){100\over{f_g(s)}}\exp{\left[-\tau(s)\right]}ds,\end{align}
from which $\zeta_{\rm{SLR}}$ can be directly computed via
Equation~(\ref{eq:zetagen}). Using the plane-parallel approximation in
conjunction with $f_g(s)=\rho_g(s)/\rho_d(s)$, we can solve
Equation~(\ref{eq:xisettle}), as was done for the well-mixed case
(Section~\ref{sec:slab}).  In general, this integral must be carried
out numerically and is moderately computationally expensive, but it 
provides an accurate treatment of the problem.  We use this numerical
result (see Figure~\ref{fig:ZetaSettle}a) to benchmark the
approximations derived below.

\subsubsection{Uniform Thin Dust Layer} 

Our first approximation considers the case where the radioactive
elements have settled into a thin layer with scale height $h\ll{H}$.  
We can write the density distribution with the limiting form 
\be\rho_d(z)=\Sigma_d\delta(z),\ee where $\delta(z)$ is the Dirac
delta function and $\Sigma_d=\Sigma_g/100$ is the total dust surface
density. With this substitution, the solution to the integral in
Equation~(\ref{formsolution}) for a gaussian disk
is\be{I}(z\gg{h};\mu)={J\Sigma_g\over4\pi\mu}\exp\left[-\frac{\tau_0}{|\mu|}\,\erf\left({z\over\sqrt{2}H}\right)\right],\ee
which is valid for $\mu\ge0$. In the limit where all SLRs are
concentrated at the midplane, $I=0$ for $\mu<0$, i.e., nothing is
emitted from above. The term in brackets is the optical depth from the
midplane to height $z$ (divided by $\mu$), and can be calculated for a
general disk. The corresponding energy density has the form 
\be\label{eq:edendelta}{c}\edense(z)=2\pi\int_0^1I(z,\mu){d\mu}={J\Sigma_g\over2}{E_1}\left[\tau_0\,\erf\left({z\over\sqrt{2}H}\right)\right],\ee
where $E_1(\tau)$ is the exponential integral of order one. In the
limit $\tau\to0$, $E_1(\tau)\to\infty$ as a natural consequence of the
$\delta$-function.  To evaluate the ionization rate near the midplane,
we now consider the density distribution to have finite thickness
\be\rho_d(z)={\Sigma_d\over2\sqrt{2}h}\qquad{\rm for}\qquad|z|\le{\sqrt{2}h},\ee 
where $\rho_d(z)=0$ for $|z|>\sqrt{2}h$. With this specification, the
specific intensity has the form
\be{I}(z;\mu)={J\over4\pi}{\Sigma_g\over2\sqrt{2}h}\int_0^\infty\Theta(z')\exp[-\tau(s)]{ds},\ee
where $\Theta(z')=1$ for $-\sqrt{2}h\le{z'}\le{\sqrt{2}h}$ and is zero
otherwise. If $z>\sqrt{2}h$, then only rays with $\mu>0$ result in
nonzero $I$. Next we define $t\equiv{z'}/\sqrt{2}H$ such that
\be\label{eq:tauset}\tau(t)={\tau_0\over|\mu|}\left|\erf\left({z\over\sqrt{2}H}\right)-\erf(t)\right|\,.\ee
For $z<\sqrt{2}h$, the specific intensity becomes 
\begin{align}{I}(z<h;\mu>0)={J\over4\pi}{\Sigma_g\over\mu}{H\over2h}\int_{-h/H}^{z/H}\exp\left[-\tau(t)\right]dt,\\
{I}(z<h;\mu<0)={J\over4\pi}{\Sigma_g\over|\mu|}{H\over2h}\int_{z/H}^{h/H}\exp\left[-\tau(t)\right]dt.\end{align}
In the limit $z\to0$, the specific intensity becomes
\be{I}(z\to0,\mu)={J\over4\pi}{\Sigma_g\over2h}{H\over|\mu|}\int_{0}^{h/H}\exp\left[-\frac{\tau_0}{|\mu|}\erf(t)\right]dt,\ee
where we have used Equation~(\ref{eq:tauset}). Next we evaluate the
specific intensity in the limit $h\ll{H}$ where we can 
approximate $\erf(t)\approx{2t}/\sqrt{\pi}$, which implies 
\be{I}(z\to0,\mu)={J\over4\pi\kappa}{\sqrt{\pi}H\over{2h}}\left\{1-\exp\left[-{\tau_0\over|\mu|}{2h\over\sqrt{\pi}H}\right]\right\}.\ee
Defining a ``dust-compactness'' parameter
\be\label{eq:lambdef}\lambda\equiv{2h\over\sqrt{\pi}H},\ee 
the corresponding energy density at $z\to0$ is given by
\begin{align}\label{eq:eden0}{c}\edense_0&={J\over\kappa}{1\over\lambda}\left\{1-\int_1^\infty{\exp\left[-\tau_0\lambda x\right]\over{x^2}}dx\right\}\nonumber\\&={J\over\kappa}{1\over\lambda}\left\{1-E_2\left[\tau_0\lambda\right]\right\}.\end{align}

Equations~(\ref{eq:edendelta}) and (\ref{eq:eden0}) give the
ionization rate $\zeta_{\rm{H_2}}$ in the limits $z\gg{h}$ and
$z\to0$, respectively.  To provide a continuous expression for
$\zeta_{\rm{H_2}}$, we smoothly connect the two limiting forms, 
\begin{align}\label{eq:zetathin}\zeta_{\rm{H_2}}(z)=(1-f(z))\left\{{J\Sigma_g\over2}{E_1}\left[\tau_0\,\erf\left({z\over\sqrt{2}H}\right)\right]\right\}\nonumber\\+f(z)\left\{{J\over\kappa}{1\over\lambda}\left\{1-E_2\left[\tau_0\lambda\right]\right\}\right\},\end{align}
where $f$ is a continuous function, e.g., $f=\exp[-z^2/2h^2]$. In the
limit $\lambda\to0$, Equation~(\ref{eq:zetathin}) provides the
solution for a disk model where all of the dust has settled to the
midplane.  In limit $\lambda\to1$, we recover the solution for the
well-mixed disk (Section~\ref{sec:slab}).

Figure~\ref{fig:ZetaSettle}~(b) shows the ionization rate due to
mm-grains (containing 85\% of the dust mass) approximated as a thin
slab from Equation~(\ref{eq:zetathin}).  In panel (c) we combine the
results in (b) with the well-mixed calculation for small grains (15\%
of all dust mass) following Section~\ref{sec:slab}, thereby producing
a ``hybrid'' model.  At large heights and large radii, the thin-slab
solution reproduces the full solution (a) quite well, though slightly
overestimates ($\sim30-60\%$) the ionization inside $R<50$~AU near the
midplane.

\subsubsection{Bidirectional Escape Approximation}\label{sec:bidirect} 

As an alternate approach, we consider only the vertical column of
emitters/absorbers, essentially assuming that radiation travels only
in the $z$-direction.  This simplification provides the solution after
evaluating only one integral in addition to $\tau$, and thus allows us
to consider more complex/non-analytic dust and gas distributions.  In
this ``bidirectional escape'' approximation, the ionization rate has
the form
\begin{align}
\zeta_{\rm{H_2}}(z)=\frac{J}{2}{\kappa\mbar\over{W_{\rm{H_2}}}}\left\{\int_{0}^{\infty}\rho_g(z)\left({100\over{f_g(z)}}\right)\exp{\left[-\tau(z)\right]}dz\right.\nonumber\\+\left.\int_{0}^{-\infty}\rho_g(z)\left({100\over f_g(z)}\right)\exp{\left[-\tau(z)\right]}dz\right\}.\end{align}
where $\tau_{\pm}$ are given by Equation~(\ref{eq:tauformal}).
Although this approximation is somewhat crude, the solution can be
readily evaluated via numerical integration, where the emissivity $J$
is given by Equation~(\ref{eq:emissivity}).
Figure~\ref{fig:ZetaSettle}~(d) compares this bidirectional
approximation to the plane-parallel calculation for the same dust
density profile.  In general, the bidirectional calculation
underestimates the ionization rate by a factor of $\sim1.5$, with
the discrepancy growing to a factor of $\sim2$ for $R\ge300$~AU. 

\subsection{Time Dependence}\label{sec:time}
\begin{figure*}\begin{centering}
\includegraphics[width=6.95in]{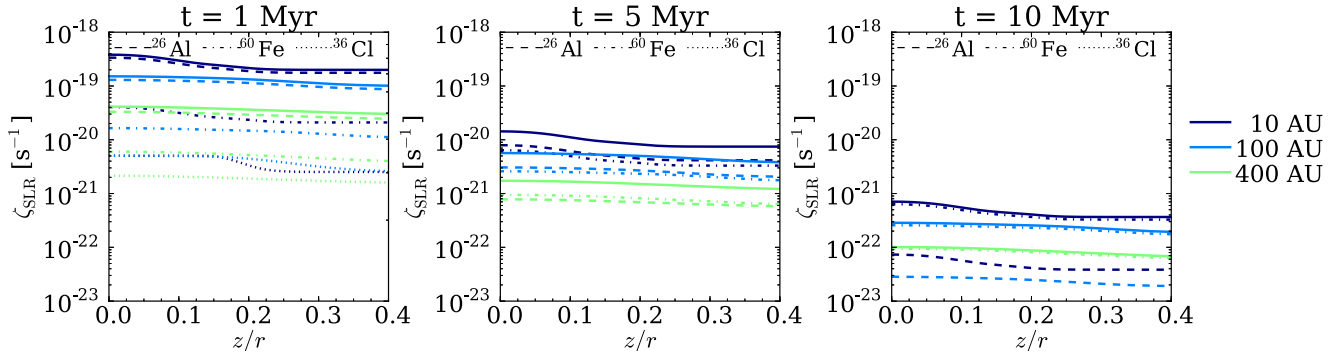}
\caption{Same as Figure~\ref{fig:ZetaAllMix}, at the indicated time  
after $t=0$ such that the ionizing reservoir has been partially
exhausted. For $t<5$~Myr, the ionization rate is set by \Al-decay.  At
later times (right, $t=10$~Myr) the longer-lived $^{60}$Fe provides the
ionization, although at a reduced rate due to its lower abundance and
longer decay time.\label{fig:timedep}}
\end{centering}\end{figure*}
Technically, the abundances of the SLRs evolve with time, with half-lives
$t_{\rm{half}}\sim1$~Myr, comparable to disk evolution timescales.  We
have calculated the ionization rates for the unsettled disk at $t=1,5$
and 10~Myr, to compare with the $t=0$ calculations from
Section~\ref{sec:radtran} (Figure~\ref{fig:ZetaAllMix}); these results
are presented in Figure~\ref{fig:timedep}.  At early times, $^{26}$Al
and $^{36}$Cl are the dominant ionizing agents; after $\sim5$~Myr,
however, the longer-lived though less abundant $^{60}$Fe determines the
ionization rate $\zeta_{\rm{SLR}}$.  At even longer times, long-lived
radionuclides such as $^{40}$K \citep[$t_{\rm{half}}\sim1.28$~Gyr;][]{umebayashi2013} 
provide the largest contribution. However, because the
ionization rate is inversely proportional to $t_{\rm{half}}$, long-lived radionuclides
only produce ionization rates of order
$\zeta_{\rm{H_2}}\lesssim10^{-22}$~s$^{-1}$.

\begin{figure}
\begin{centering}
\includegraphics[width=2.66in]{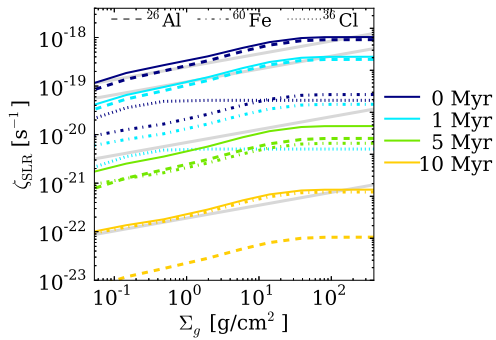}
\caption{Ionization rate per H$_2$ in the midplane as a function of
disk surface density at the indicated times. We fit these results
using a simple power-law versus surface density, normalized with to a
time-dependent constant (gray lines), see Equation~(\ref{eq:approxfit}).}\label{fig:sigion}
\end{centering}
\end{figure}

Figure~\ref{fig:sigion} plots the ionization rate $\zeta_{\rm{SLR}}$
calculated at the disk midplane ($z=0$) versus vertical disk surface
density at several times.  We have fitted power-laws
(shown in gray in Figure~\ref{fig:sigion}) to the ionization rates
from all three SLR species to facilitate their use. For the well-mixed
disk, the rate $\zeta_{\rm{H_2}}$, as a function of disk surface
density and time, is given by
\be\label{eq:approxfit}\zeta_{\rm{H_2}}(r)=\left(2.5\times10^{-19}~{\rm{s}}^{-1}\right)
\left(1\over2\right)^{1.04t}\left({\Sigma(r)\over{{\rm{g~cm^{-2}}}}}\right)^{0.27},\ee 
where time, $t$, is given in Myr.

\section{Discussion}\label{sec:discussion}

This work has made two simplifying assumptions that warrant further
consideration.  First, we have used a single opacity to simplify the
calculations.  In reality, the photons and particles will have their
energies degraded as they collide with the gas. For Compton-scattered
photons, the change in $\gamma$-ray energy for $E_\gamma<m_ec^2$ is approximately $\Delta{E}\propto{E_\gamma}^2$ \citep{rybicki1979}.  This energy degradation results in an energy spectrum of the form $I(E_\gamma)\approx{E_\gamma}~dN/dE\approx{E_\gamma}/\Delta{E}\propto1/E_\gamma$, from which we may compute
a weighted opacity for photons that lose all of their energy
through scattering in the disk,
\be\langle\kappa\rangle\equiv{\int_0^{E_i}\kappa(E_\gamma)I{dE}\over\int_0^{E_i}{I}{dE}}\,.\ee
At sufficiently low energies, below $E_\gamma<30$~keV, the dominant
absorption mechanism becomes photoabsorption, at which point the
photon has lost nearly all of its initial energy $E_i\sim1$~MeV. 
The Compton-weighted opacity for a 1.808~MeV photon is
$\langle\kappa\rangle\sim0.19~{\rm{cm^{2}~g^{-1}}}$, larger than its 
initial cross section, $\kappa_i\sim0.08~{\rm{cm^{2}~g^{-1}}}$, which
increases the derived ionization rate by factors of 4\% (12\%) at 
$r=10$~AU (400~AU). 

Another major simplification comes from our treatment of the positrons
in \Al\ decay.  Currently we assume that all energetic particles are
emitted locally and can then escape.  According to
\citet{umebayashi2013}, the positron will first lose its energy to
primarily collisional ionization until it comes to rest, at which
point it will annihilate with an electron and produce two 0.511~MeV
$\gamma$-rays. However, the cross section for absorption of the
$\gamma$-rays is less than that of the original positrons. For disk
regions where positrons escape in our simplified treatment, these
$\gamma$-rays escape even more readily, and thus make smaller
contributions to the ionization rate.

\section{Conclusion}\label{sec:conclusion}

This paper carries out radiative transfer calculations for the decay
products of short-lived radionuclides in circumstellar disks. These
SLRs provide an important contribution to the ionization rates, which
in turn affect disk chemistry and physics. We
provide simple analytic expressions for the ionization rates due to
SLR decay. For well-mixed disks, the ionization rate
$\zeta_{\rm{SLR}}(r,z)$ can be found analytically and is given by
Equation (\ref{eq:zeta_mix}). The radial dependence is controlled by
the surface density profile, which determines the optical depths
through Equations (\ref{eq:tauformal}) and (\ref{eq:tauzero}).
Complications arise as disks evolve, including dust settling and a
decrease in the SLR abundances. We provide two approximations for
disks with settled dust layers: The first treatment considers the dust
as a thin uniform layer; the ionization rate is given by Equation
(\ref{eq:zetathin}), where the dust-compactness parameter $\lambda$
determines the degree of settling (Equation (\ref{eq:lambdef})). This
approximation scheme is accurate to tens of percent, and becomes exact
for a well-mixed disk. For completeness, we develop a simpler
approximation that considers radiation propagation in the vertical
directions only (Section \ref{sec:bidirect}), as is commonly done for
calculations of CR ionization rates. Finally, we provide a
fit to the midplane ionization rates as a function of surface density
and time (Equation~(\ref{eq:approxfit})).  While this function can be 
applied over the entire vertical structure, it formally overestimates 
$\zeta_{\rm{H_2}}^{\rm{SLR}}$ at the disk surface. In this regime, however, other
ionization sources (e.g., stellar X-ray photoionization of H$_2$) will
dominate, so that our approximation remains satisfactory.

A full treatment of this problem requires Monte Carlo or other
numerical methods, including energy losses, angle-dependent
scattering, energy-dependent radiative transfer, and more
sophisticated density distributions. Although these generalizations
should be incorporated in future work, the analytic expressions
derived herein provide useful and accurate estimates for the
ionization rates due to SLR decay, and can thus be used in a wide
variety of physical and chemical disk models.

\acknowledgments{This work was supported by NSF grant AST-1008800.}

\end{document}